\documentclass[journal]{IEEEtran}
\ifCLASSINFOpdf
\else
\fi
\usepackage{graphicx}
\usepackage{cite}
\usepackage[ampersand]{easylist}
\usepackage{rotating}
\usepackage{amsmath,amssymb,amsfonts}
\usepackage{algorithmic}
\usepackage{textcomp}
\usepackage{multirow}
\usepackage{array, multirow, makecell}
\usepackage{array}
\usepackage{booktabs}

\begin{document}
%
\title{Quality of Experience for Streaming Services: Measurements, Challenges and Insights}
%
%
%

\author{Khadija~Bouraqia$^\dag$,~\IEEEmembership{Student Member, IEEE,}
        Essaid~Sabir$^\dag$,~\IEEEmembership{Senior Member, IEEE,}
        Mohamed~Sadik$^\dag$,~\IEEEmembership{Member, IEEE,}
        and Latif Ladid$^\ddag$,~\IEEEmembership{Senior Member,IEEE}\\
        $^\dag$NEST Research Group, ENSEM, Hassan II University of CAsablanca, Morocco\\
        $^\ddag$University of Luxembourg, Luxembourg
}

%
%

\markboth
{K. Bouraqia \MakeLowercase{\textit{et al.}}: Quality of Experience for Streaming Services}
{arXiv Preprint, December~2019}

%



\maketitle
\begin{abstract}
Over the last few years, the evolution of network and user handsets' technologies, have challenged the telecom industry and the Internet ecosystem. Especially, the unprecedented progress of multimedia streaming services like YouTube, Vimeo and DailyMotion resulted in an impressive demand growth and a significant need of Quality of Service (QoS) (e.g., high data rate, low latency/jitter, etc.). Mainly, numerous difficulties are to be considered while delivering a specific service, such as a strict QoS, human-centric features, massive number of devices, heterogeneous devices and networks, and uncontrollable environments. Thenceforth, the concept of Quality of Experience (QoE) is gaining visibility, and tremendous research efforts have been spent on improving and/or delivering reliable and added-value services, at a high user experience. In this paper, we present the importance of QoE in wireless and mobile networks (4G, 5G, and beyond), by providing standard definitions and the most important measurement methods developed. Moreover, we exhibit notable enhancements and controlling approaches proposed by researchers to meet the user expectation in terms of service experience. 
\end{abstract}

\begin{IEEEkeywords}
Quality of Experience(QoE); Quality of Service (QoS); QoE Measurements; QoE Enhancements; 4G/5G/B5G; D2D; M2M; EDGE Computing; Content Caching.
\end{IEEEkeywords}

%
\IEEEpeerreviewmaketitle

\section{Introduction}
%
%
%
%
Until recently the quality of service (QoS) \cite{cheng2016statistical} provided has been evaluated from a technical perspective to determine network performance, through measuring several factors (i.e., throughput, available bandwidth, delay, error probability, jitter, packet loss, etc.). Nonetheless, for many services like video streaming, QoS cannot capture the influence of the network fluctuation on the user experience \cite{seufert2019considering}. 
 In 1994, and according to the International Telecommunication Union (ITU) recommendation, ITU-T Rec.E.800, \cite{recommendation1994800} the quality of service was defined as:
 \begin{center}
 \textit{``Collective effect of service performance which determines the degree of satisfaction of a user of the service''}
 \end{center}
\noindent Markaki redefined it \cite{4394198} as the, 
 
\begin{center}
\textit{``Capability of a network to provide better service to selected network traffic ... described by the following parameters: delay and jitter, loss probability, reliability, throughput and delivery time''}
\end{center}   
As we notice in the second definition, user satisfaction is not considered anymore. Giving that many service providers are competing for more costumers; a new notion has emerged Quality of Experience (QoE), used instead of QoS to enhance the service and get the consumer's feedback on a specific service (e.g., network). The QoE is related to both objective QoS (i.e., objective metrics depict the influence of the network and application performance on the user) and subjective \cite{zhao2016qoe} (i.e., the individual user experience obtained from expectation, emotional state, feeling, preference, etc.). 
%
%
In other words, it is an evaluation of individuals' experience when interacting with technology and business entities in a particular context \cite{laghari2012toward} to provide satisfaction to the end-user. \\
Here we introduce some definitions of this new concept by starting with the most used definition for QoE:
\begin{center}
\textit{``Overall acceptability of an application or service as perceived subjectively by the end-user ... includes the complete end-to-end system effects ... maybe influenced by user expectations and context.''}
\end{center}
by ITU-T SG 12 in 2007 \cite{hands2008standardization} but it does not clarify what the QoE is about and how it could be measured. 
Based on ITU-T SG 12 2007 and Dagstuhl seminar 2009 \cite{seminar2009quality} a new influencing factor, context, was added as follows:
\begin{center}
\textit{``Degree of delight or annoyance of the user of an application or service as perceived subjectively includes the complete end-to-end system effects ... maybe influenced by user state, content and context.''}
\end{center}
 Afterward, a better definition was also proposed in the Dagstuhl Seminar \cite{seminar2009quality}:
\begin{center}
 \textit{``Describes the degree of delight of the user of a service, influenced by content, network, device, application, user expectations, and goals, and context of use.'' } 
\end{center}
\noindent The last one as far as we know, is considered as a working definition of QoE is \cite{brunnstrom2013qualinet}:
\begin{center}
 \textit{``QoE is the degree of delight or annoyance of the user of an application or service. It results from the fulfillment of his or her expectations with respect to the utility and/or enjoyment of the application or service in the light of the user's personality and current state.''}
\end{center}
\noindent We conclude that from all the definitions mentioned
above, there is no practical or exact definition to explain the substance of the QoE, how to measure it, or what it impacts on the users' expectations. However, these definitions give a broad understanding of the QoE, which offers an excellent opportunity to research and explore it in depth. \\
\noindent The rest of this paper is structured as follows. We provide an overview of the influencing factors on the users' experience in section II. We introduce different models and approaches used to measure the QoE in section III. Then, in section IV, we discuss controlling methods proposed by various researchers to improve the QoE, section V exhibits the challenges and enhancements aiming to bring the content closer to the end-user. In section VI, we discuss some recent technologies and hot problems related to QoE. Finally, a few concluding observations are drawn in Section VII. 

 

\section{Factors influencing the Quality of Experience}
\begin{figure}[!ht]
\centering
 \includegraphics[scale=0.2]{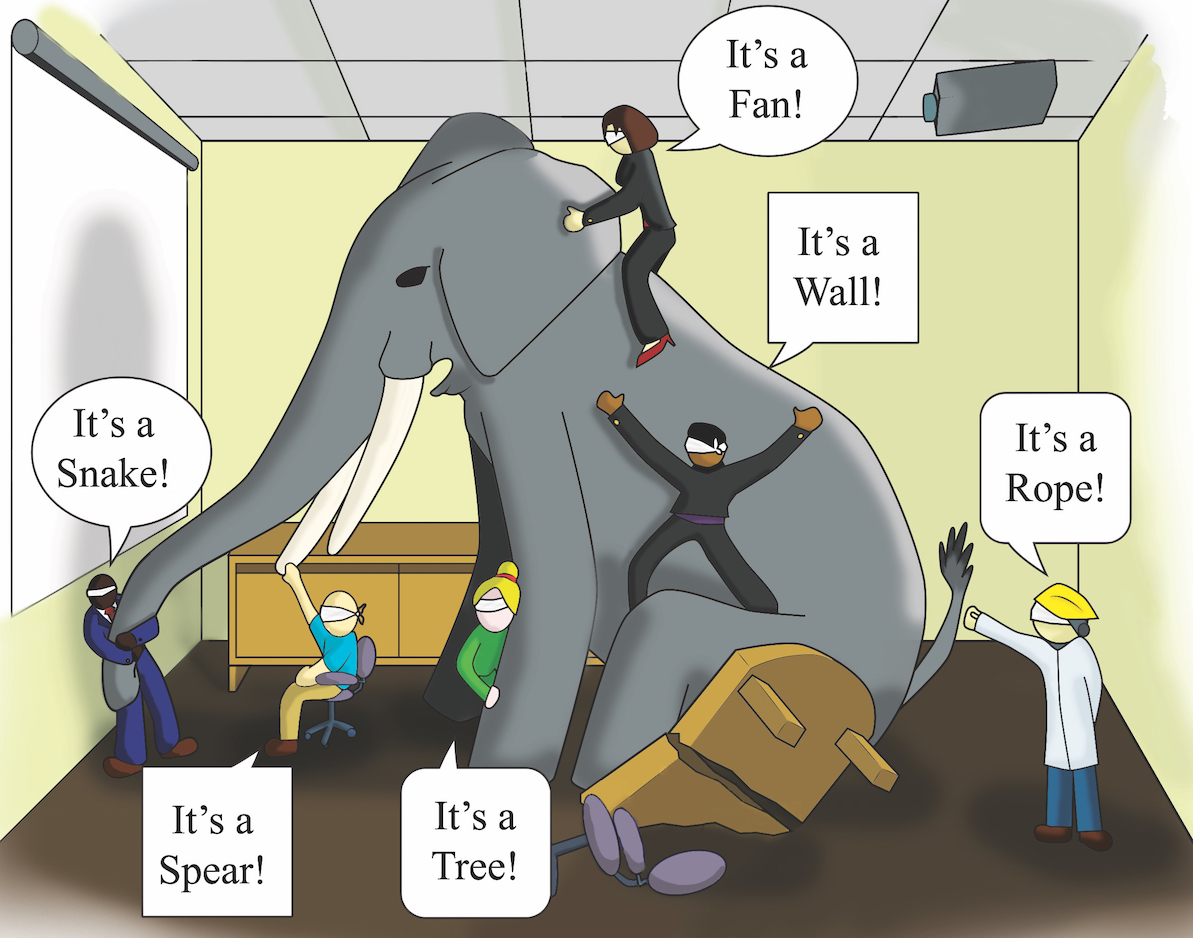} 
\caption{Challenging subjective evaluation from different perspectives.}
\label{perspective}
\end{figure}

Since the QoE is still a new concept, content providers, service and network providers, in addition to researchers are facing new challenges related to delivering, measuring, and controlling QoE. Then, investigating and analyzing the QoE influencing parameters (IFs) \cite{barakovic2019quality} is a first step to go. It is hard to predict the QoE because of its subjective nature, see Figure \ref{perspective}. Therefore, in order to evaluate the overall service quality, factors that influence the users' perception should be determined beforehand \cite{mellouk2013quality}. Qualinet \cite{brunnstrom2013qualinet} has defined IFs of the QoE as follows:
 \begin{center}
 \textit{''Any characteristic of a user, system, service, application, or context whose actual state or setting may have influence on the Quality of Experience for the user.''}\\
 \end{center}
 
The IFs could interrelate, thus they should not be classified as isolated entities. From this perspective, they are classified into three categories:

\begin{itemize}
\vspace{0.2cm}
\item \textbf{Human-related Influencing Factor}: any variant or invariant property or characteristic of a human user. The characteristic can describe the demographic and socio-economic background, the physical and mental constitution, or the user's emotional state. \vspace{0.2cm}
\item \textbf{System-related Influencing Factors}: properties and characteristics that define the technically generated quality of a service or an application. They are associated to media capture, transmission, coding, storage, rendering, and reproduction/display, also to the communication of information itself from content production to the user. \vspace{0.2cm}
\item \textbf{Context-related Influencing Factors}: are factors that embrace any situation property to describe the user's environment, in terms of physical (location and space, including movements within and transitions), temporal, social (people present or involved in the experience), economic (Costs, subscription type, or brand of the service/system), task, and technical characteristics These factors can occur on different levels. 
\end{itemize}
\begin{figure}[!htb]
\centering
 \includegraphics[width=0.5\textwidth]{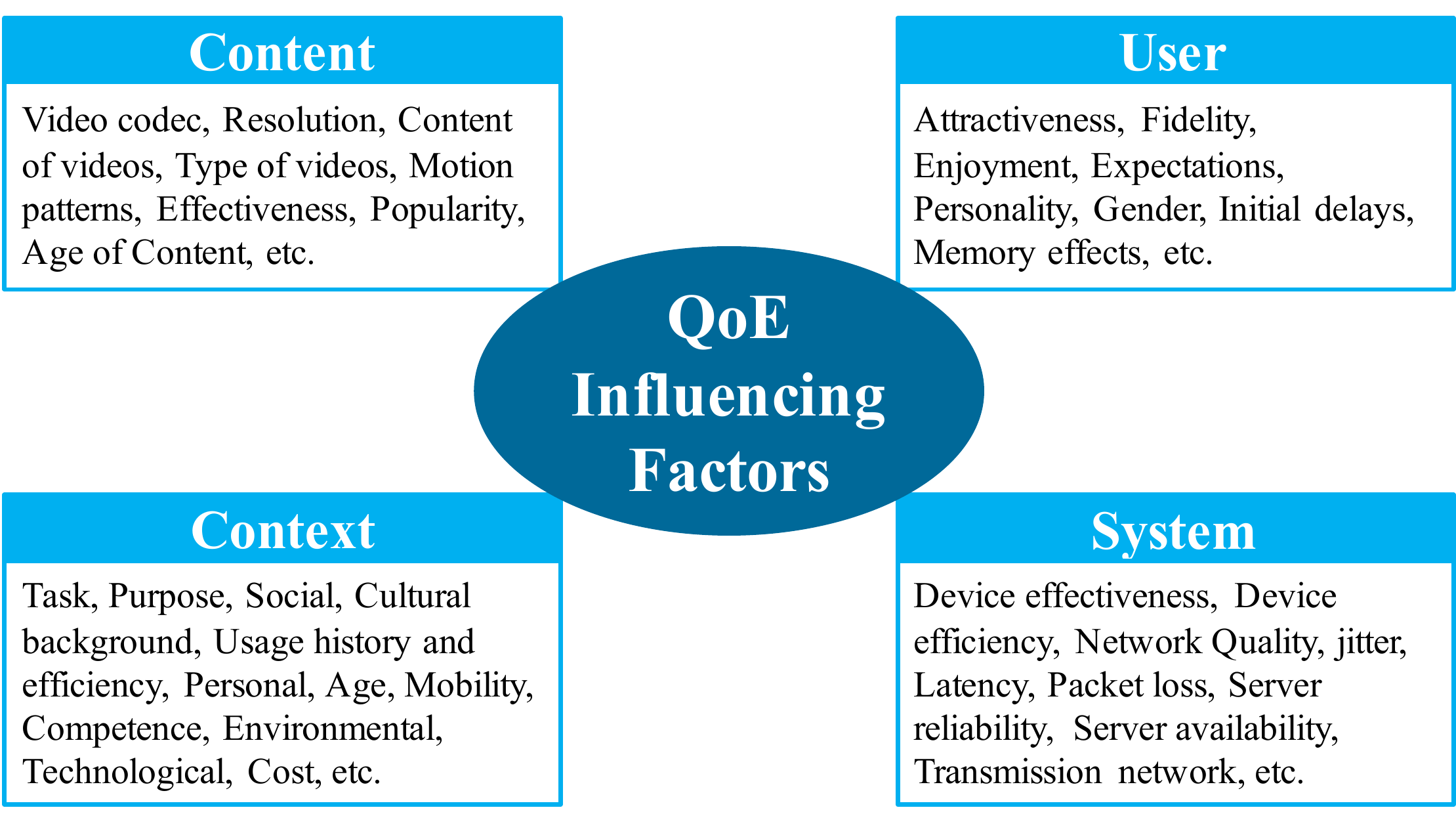} 
\caption{Different Factors influencing QoE.}
\label{IF}
\end{figure}
\vspace{0.2cm}
\noindent In addition to the three previous IFs (i.e., context level, system level and user level), Juluri \emph{et al. } \cite{juluri2015measurement} introduced a fourth IF for video delivery, see Figure \ref{IF}:
\begin{itemize}
\vspace{0.2cm}
\item \textbf{Content-related Influencing Factors}
the information regarding the offered content by the service or application under study. They are associated, in the case of video, with video format, encoding rate, resolution, duration, motion patterns, type and contents of the video, etc.\\
\end{itemize} 
\noindent Several works provided other external factors. Like the importance of the application, user's terminal hardware, and mobility \cite{mellouk2013quality}. Also, five standards of video quality metrics (i.e., the join time, the buffer ratio, the rate of buffer events, average bit-rate, and rendering quality) were presented in\cite{dobrian2013understanding}. As well as the prefetching process, source coding \cite{xu2014flow} and the effect of packet reordering \cite{usman2017no, jaiswal2007measurement} studied in \cite{laghari2018measuring}. 
\\
In another perspective, a comparison of the influence of some metrics the packet loss and bandwidth have a significant impact than the jitter and delay \cite{vega2018resilience}. 
In short, it is worth noting that there are specific IFs relevant for different types of services and applications.
\section{Measurements Approaches}
To consider the user satisfaction in the context of real-time video streaming applications, QoS is no longer sufficient to evaluate the quality. Therefore, researches have been conducted to assess the QoE \cite{yang2018survey}. 
In this section, we will address the developed techniques to measure the QoE \cite{ITUG.1011}. 
\\Whether using subjective or objective methods or combine both are discussed in \cite{mellouk2013quality} as follows: ``Subjective methods are conducted to obtain information on the quality of multimedia services using opinion scores, while objective methods are used to estimate the network performance using models that approximate the results of subjective quality evaluation.''
\subsection{Subjective Assessment}
\vspace{0.2cm}
In \cite{matulin2014subjective}, subjective assessment is considered as the most accurate approach to measure the QoE perceived by the end-user. This method gathers human observers in a laboratory to evaluate sequences of a video and then scores depending on their point of view and their perception, the average of the values obtained for each test sequence is known as the Mean Opinion Score (MOS)\cite{ITU}; MOS is often used to quantify these factors. Commonly rated on a five-point discrete scale as follows \textit{[1:bad, 2:poor; 3: fair; 4:good; 5:excellent]}. Although MOS is the most known precise assessment, it slows scoring due thinking and interpretation, as well people are limited by finite memory and cannot capture users perception over time. 
In addition, in a recent research \cite{nilsson2018suitability} authors have studied the impact of considering young student (9-17 years old) as viewers to evaluate the quality of videos (MOS) subjectively. The results suggested that they are suitable and can notice different quality issues to the adults. However, more studies should be performed. 
\\
To conduct a subjective quality test, to evaluate a video quality \cite{rec2009bt}, we introduce some of the widely known standard methods as follows:
\begin{itemize}
\vspace{0.2cm}
\item Double Stimulus Continuous Quality Scale (DSCQS)\cite{BT.500-11}: The evaluator is presented twice by reference and the processed video sequence in alternative fashion, upon termination of the video he is asked to rate its quality at a scale of 0 (lowest value)-100 (highest value) then the difference of the video assessment value is calculated. In the case of a small value, the quality of the presented video is close to the reference video else the quality is low. For a large number of video scenes, DSCQS needs a very long time to implement quality assessments. \vspace{0.2cm}
\item Single Stimulus Continuous Quality Evaluation (SSCQE) \cite{BT.500-11} ITU-R recommendation:
The user votes the quality of a continuous video usually of 20 to 30 minutes. This method allows observing the variation of the quality over time by calculating the average quality evaluation of the subjects, SSCQE requires well-trained observers to attain stable assessment results. \vspace{0.2cm}

\item Absolute Category Rating (ACR)\cite{ITU-TRec.P.910}:
ACR is recognized as a single stimulus method. The video is watched for about 10 seconds, and during the next interval up to 10 seconds, the subjects evaluate the video by the five-grade quality scale expressed as MOS.  \vspace{0.2cm}

\item Absolute Category Rating-Hidden Reference (ACR-HR) \cite{ITU-TRec.P.910}: This approach is similar to ACR. Except that the reference version of each
shown distorted test sequence is also displayed to the participants. Afterward, they give their scores in the form of MOS, and a final quality evaluation is computed using a differential quality score. \vspace{0.2cm}

\item Pair Comparison \cite{ITU-TRec.P.910}:
Pair of videos are presented to the subjects to be compared and then evaluated (i.e., which one of the pairs has superior quality). The results vary depending on, which one was shown first, as the assessments take longer time than the ACR method. \\
\end{itemize}

\noindent Other standards, such as Simultaneous Double Stimulus for Continuous Evaluation, Subjective Assessment Methodology for Video Quality, Degradation Category Rating or DoubleStimulus Impairment Scale and Comparison Category Rating, are discussed in \cite{gu2017no}. 
\newline
\\
\noindent Subjective Assessments are very expansive in terms of human resources, cost and time consumption. However, such technique cannot be used as an automatic measurement or monitoring for real-time applications like video streaming. Fortunately, there exists another subjective evaluation form of QoE, that enables new potentialities to conduct web-based tests. It is more flexible, offers a diverse population as participants and is cost and time effective. Besides, it creates a realistic test environment, named Crowdsourcing \cite{hobfeld2014survey, naderi2018motivation}. Here, we cite some platforms and web-based frameworks:
\begin{description}
\vspace{0.2cm}
\item [-] \hspace{-0.7cm}\textbf{Aggregator platforms} (e.g., Crowdflower, Crowdsource): These platforms often delegate the task to different channels, that provide workers. Such a system focuses on a limited set of predefined tasks only. Meanwhile, it might suffer from a significant drawback as some aspects of the experiment, might not be directly controllable;
 
\item [-] \hspace{-0.7cm}\textbf{Specialized platforms} (e.g.,  Microtask, TaskRabbit): This platform focuses on a limited set of tasks or a specific workers class, as it maintains their workers;

\item [-] \hspace{-0.7cm}\textbf{Crowd providers} (e.g., Amazon, Mechanical Turk, Microworkers, TaskCN): Acknowledged as the most flexible type, a self-organizing service, maintains a large work crowd and offers unfiltered access to the recruited participants; 
 
\item [-] \hspace{-0.7cm}\textbf{Quadrant of Euphoria}: Permits for a pairwise comparison of two different stimuli, so the worker could judge which of the two stimuli has a higher QoE. A test uncovers fake users and rejects them, but at the cost of exposing reliable users also to rejection.
\vspace{0.2cm}
\end{description}
On the other hand, an underdeveloped crowdsourcing system is proposed \cite{lin2016crowdsourcing}, to evaluate the QoE of video on demand streaming. This system is different from other crowdsourcing platforms as it can monitor network traffic and the bandwidth, as well measure the central processing unit (CPU) usage, Random Access Memory (RAM) utilization, times video freezes and MOS (i.e., users fill a questionnaire). It proved to be about a 100\% accurate in High Definition display resolution (HD) and about 81 to 91\% in other qualities as their test shows. 
\\

\noindent Most of these Crowdsourcing techniques have only allowed testers to conduct the test on their computers or laptops. However, Seufert \emph{et al. }\cite{seufert2018app} introduced a new application "CroQoE". It runs on mobile devices to evaluate the QoE of streaming videos, connected to a Linux back-end server to dynamically prepare and evaluate the test. Also, they allowed users to choose the content of videos they would like to watch. 
 The results proved that this added feature (i.e., choosing the content) could slightly enhance the QoE ratings. Still, they utilized, in their tests, only high definition videos with a duration of fewer minutes. 
Crowdsourcing technique has some drawbacks, as there is a little control over the environment, which may give the participants a chance to cheat in order to increase their income. Also, as stated in \cite{egger2017crowdsourcing}, crowd diversity and expectations, the context, type of equipment (workers typically use their own devices and could differ regarding hardware, software, and connectivity) and the duration and design of the test (small duration will encourage the workers while long duration may be unreliable) impact the QoE assessment. 
\subsection{Objective Assessment} 
\vspace{0.2cm}
A considerable number of objective quality measurements have been developed using mathematical formulas or algorithms to estimate the QoE based on QoS metrics ( parameters collected from the network). Depending on the accessibility of the source signal, they are organized into three approaches:
\begin{itemize}
\vspace{0.2cm}
\item \textbf{Full reference (FR)}: a reference video is compared frame-by-frame (e.g., color processing, spatial and temporal features, contrast features) with a distorted video sequence to obtain
the quality (commonly used in lab-testing environments, e.g., ITU-T J.247). \vspace{0.2cm}
\item \textbf{Reduced reference (RR)}: Only some features of the reference signal are extracted and employed to evaluate the quality of the distorted signal (e.g., ITU-T J.246). \vspace{0.2cm}
\item \textbf{No reference (NR)}: The reference video is inessential
 while evaluating the distorted video
sequences Quality. (commonly used for real-time quality assessment of videos, e.g., ITU-T P.1201). \\
\end{itemize} 
Some of the most known objective quality assessment approaches are Peak Signal to Noise Ratio (PSNR), Structural Similarity Metric (SSIM) \cite{wang2004video}, Multi-Scale Structural SiMilarity \cite{wang2003multiscale}, SSIMplus \cite{rehman2015display}(supports cross frame rate and cross resolution), Video Quality Model (VQM)\cite{pinson2004new}, and
Natural Image Quality Evaluator (NIQE) \cite{mittal2013making}. Despite that these models outperform PSNR, 
most researchers commonly use PSNR \cite{itupsnr}, the logarithmic ratio between the maximum value of a signal and the background noise, due to its simplicity to assess video quality. However, it cannot be appropriate to be used in a real-time mechanism. 
\begin{table}[!h]
\caption{Mean Opinion Score}
\label{MOS PSNR}
\begin{center}
\begin{tabular}{|c||c||c||c||c|}
\hline
\textbf{\normalsize{MOS}} & \textbf{\normalsize{Quality}} & \textbf{\normalsize{Perception}} & \textbf{\normalsize{PSNR (dB)}}\\
\hline
 5 & Excellent & Imperceptible & $\succ 37$\\
\hline
4 & good & Perceptible & 31 - 37 \\
\hline
3 & Fair & Slightly annoying & 25 - 31 \\
\hline
2 & Poor & Annoying & 20 - 25 \\
\hline
1 & Bad & Very annoying & $\prec 20$\\
\hline
\end{tabular}
\end{center}
\end{table}
A heuristic mapping of PSNR to MOS (see Table I) exists though, the research in \cite{4550695} revealed that the correlation between the PSNR and subjective quality would be decreased if the codec type of the video content changes unless otherwise. PSNR is a qualified indicator of video quality. Here we exhibit few PSNR to MOS mapping models:\\
\begin{itemize}
\item The relation between PSNR and MOS for time-variant of video streams quality on mobile terminals~\cite{nemethova2006psnr}:
\begin{equation}
PSNR(n) = 10\cdot\log \left(\frac{255^{2}}{MSE(n)}\right)
\label{psnr1}
\end{equation}
where $MSE(n)$ is defined as follows:
\begin{equation}
MSE(n) = \frac{\sum\limits_{i=1}^{N}\sum\limits_{j=1}^{M}\sum\limits_{c=1}^{C}\big(F^c_{n}(i, j)-R^c_{n}(i, j)\big)^2}{N\cdot M\cdot C} 
\label{mse}
\end{equation}
Further, $\widehat{MOS}_{PSNR}(n)$ is captured using a linear law
\begin{equation}
\widehat{MOS}_{PSNR}(n)= a\cdot PSNR(n) + b
\label{mos1}
\end{equation}
with
\begin{displaymath}
 \textit{a}= \frac{c_{MOS, PSNR}}{\sigma^2_{PSNR}} \quad\mbox{and}
\quad
 \textit{b} = \mu_{MOS} - a. \mu_{PSNR}
\end{displaymath}
where $MSE(n)$ denotes the mean square error of the $n$-${th}$ frame $F_{n}$ compared to the frame $R_{n}$ of the reference sequence. 
\\
$i$ and $j$ address particular pixel values within the frame. 
\\ $C$ is the number of the color components and $c$ is an index to address them. 
\\ 
$c^{MOS, PSNR}$ represents the sample co-variance between the PSNR(n), and the MOS(n). 
\\
 $\mu_{PSNR}$ and $\mu_{MOS}$ are the sample means of PSNR respectively MOS. 
\\
$\sigma^{2}$ PSNR is the sample variance of PSNR. 
\\ $a$ and $b$ are respectively the scaling and the shift factors. \\
\item The PSNR-MOS nonlinear mapping model on the wireless mobile network for video services as follows \cite{2014qoe}:

\begin{equation}
PSNR = 10 \cdot  \log\left(\frac{255^{2}}{\frac{a}{\exp\left(\frac{R_{p}}{b}\right) - 1}+ \beta \cdot  PLR}\right)
\end{equation}
\\$a$ and $b$ are model parameters associated with measured data, $R_{p}$ transmitted rate of the the video service and $PLR$ is the packet loss rate. 
\begin{equation}
 MOS=\begin{cases}
  \displaystyle 1, & \scriptstyle PSNR \leq20.\\
  \displaystyle \alpha\cdot th(\xi\cdot \scriptstyle PSNR \displaystyle-\beta)+\gamma, & \scriptstyle 20<PSNR<50.\\
  \displaystyle 5, & \scriptstyle \geq50.
 \end{cases}
 \label{mos}
\end{equation}
$\alpha$, $\beta$, $\xi$ and $\gamma$ are parameters that vary with the content and structure of the video sequences. \\
\item PSNR to MOS mapping using an S-type (sigmoidal) mapping function\cite{2012mapping}:
\begin{equation}
MOS =\frac{1}{\alpha + \exp\left(\beta  (\gamma - \textit{PSNR}) \right)}+ \lambda
\label{mosmapping}
\end{equation}
\end{itemize}
$\alpha$, $ \beta$, $ \gamma$ and $\lambda$ are related parameters that can be determined through many experiments. 
Moreover, authors in \cite{he2018qoe}, based on the article \cite{wang2017crowdsourcing}, have evaluated a relationship between MOS and the bit-rate as follows:
\begin{equation}
 MOS_{Video}=\begin{cases}
  \displaystyle 0.5, & R <5\mbox{kbps}. \\
  \displaystyle \alpha \cdot \log(\beta\cdot R), & 5\mbox{kbps}\leq R <250\mbox{kbps}\\
  \displaystyle 4. 0, & R \geq250\mbox{kbps}. 
 \end{cases}
 \label{mos2}
\end{equation}
where $R$ is the bit-rate, $\alpha$ and $\beta$ the parameters obtained from the upper and lower limit of MOS values. Based on the paper \cite{wang2016game} $\alpha = 2.3473$ and $\beta= 0.2667$. After presenting PSNR; Other Frameworks were proposed to measure and predict future QoE collapses, such as:\\
\begin{itemize} 
\item The bit-rate switching mechanism is executed at the users' side in a wireless network, to elevate the quality of the user and determine the QoE metrics. 
 Xu \emph{et al. } propose\cite{xu2014analytical} a framework for dynamic adaptive streaming, that, given the bit-rate switching logic, computes the starvation probability of playout buffer, continuous playback time and mean video quality. It can be used to predict the QoE metrics of dynamic adaptive streaming.\\
\item YoMoapp \cite{wamser2015yomoapp}, a passive android application was employed in a field study of mobile YouTube video conducted in \cite{seufert2016application} to monitor the application-level key performance indicators (i.e., buffer and the video resolution) of YouTube in the user's mobile device, this monitoring application works on JavaScript which might indicate some errors however it is accurate by approximately $1$ second. \\
\item Pytomo \cite{juluri2011pytomo} evaluates the playback of a played YouTube \cite{bouraqia2017youtube} video as experienced by users. It collects the download statistics such as the ping, the downloaded playback statistics, number of stalling event and the total buffer duration, then estimates the playout buffer level. Moreover, Pytomo allows the study of the impact of the DNS resolution. This tool could be YoMo complementary. However, it is not feasible, due to the need to access the user's device. \\
\item An application for mobile service \cite{chihani2014user} was proposed to measure the QoE directly from the user's device, in order to transmit the results to the service provider while preserving the user's privacy. \\
\item QMON\cite{eckert2013qoe} is a network-based approach that monitors and estimate the QoE of the transmitted video streaming. It focuses on the occurrence and the duration of playback stalls, also it supports a wide range of encoding (MP4, FLV and WebM). 
 The study confirmed that streaming parameters (i.e.,  stalling times, times on quality layers) are the best appropriate for QoE monitoring, to ensure an accurate developed model to estimate
QoE. \\
\item The authors in \cite{xu2014flow} studied the quality of streaming from the aspect of flow dynamic. They developed an analytical framework that computes the QoE metrics like dynamics of playout buffer, scheduling duration, and the video playback variation, in a streaming service over wireless networks. The framework is assumed to anticipate precisely the distribution of prefetching delay and the probability of generating a function of the buffer starvation. The obtained result proved that the flow dynamics has more influence on QoE metrics. 
Also, it is assumed to be suitable in some scenarios like hyper-exponential video length distribution, heterogeneous channel gains, mixed data, and streaming flow. \\
\item Network operators may handle long and short views with different priorities. Thus \cite{xu2016modeling} build a model on starvation behavior in a bandwidth sharing wireless network by using a two-dimensional continuous time Markov process and ordinary differential equations to determine that progressive downloading increases, considerably, the starvation probability. Further, they observed based on their result, that the history of time-independent streaming traffic pattern can predict future traffic, and that the viewing time follows a hyper-exponential distribution which is validated to be more accurate than some existing models (i.e., exponential, Pareto distribution). \\
\item The paper \cite{liu2014software} proposes a real-time video QoE software assessment system. It evaluates the error of network in the part of video transmission, by testing the value of the service quality, the quality of transmission, the encoded videos in various contents and sizes. The authors indicate that this platform is deployable on a real network. \\
\item A QoE Index for Streaming Video (SQI) model was proposed by Duanmu \emph{et al. } \cite{duanmu2017quality} to predict the QoE instantly. To build their model, they have started by constructing a video database (effect of initial buffering, stalling, video compression), then investigate the interactions between video quality and playback stalling. The SQI seems to be ideal for the optimization of media streaming systems as well; it is simple in expression and effective. However, it does not support reporting function on the degradation of QoE and has limited monitoring parameters. \\
\item YOUQMON \cite{casas2013youqmon} estimates the QoE of YouTube videos in real time in 3G networks. It combines passive traffic analysis and a QoE model to detect stalling events and project them into MOS. Each minute monitoring system computes the number of stalling as the fraction of stalling of every detected video, as well it supports two video formats used by YouTube, AdobeFlash, and Moving Picture Experts Group (MPEG4). The results appear to be accurate, similar to  MOS values and indicate the potentiality of the performance of this system. Still, it cannot identify the point of the network that impacts the quality. \\
\item The QoE Doctor tool \cite{chen2014qoe} is an Android tool that can analyze across different layers (application, transport, and network), from the app user interface (UI) to the network. The tool employs a UI automation tool to duplicate user behavior and to measure the user-perceived latency (i.e., identify changes on the screen), mobile data consumption, and network energy consumption. QoE Doctor can quantify the factors that impact the app QoE and detect the causes of QoE degradation, although it is unfit to supervise or control the mobile network, the component responsible for detecting UI changes has to be adjusted for each specific app.\\
\item Zabrovskiy \emph{et al. } \cite{zabrovskiy2017advise} presented AdViSE, an Adaptive Video Streaming Evaluation framework of web-based media players, and adaptation algorithms. It supports different media formats, various networking parameters and implementations of adaptation algorithms. 
AdViSE contains a set of QoS and QoE metrics gathered and assessed during the adaptive streaming assessment evaluation as well as a log of segment requests, applied to generate the impaired media sequences employed for subjective evaluation. 
Still, they do not provide a source code level analysis of familiar Dynamic Adaptive Streaming over HTTP (DASH) players and support for popular commercial streaming players. 
 In \cite{timmerer2018automated}, same authors proposed an end-to-end QoE evaluation to collect and analyze objectively (AdViSE) and subjectively (Web-based subjective evaluation platform (WESP) \cite{rainer2013web}) the streaming performance metrics (e.g., start-up time, stalls, quality). The framework is flexible and can also determine when players/algorithms compete for bandwidth in different configurations although it does not consider Content Delivery Network (CDNs), Software-Defined Networking (SDN), nor 5G networks. \\
\item VideoNOC \cite{mangla2018videonoc} is a video QoE monitoring prototype platform for Mobile Network Operators, considering video QoE metrics (e.g., bit-rate, rebuffering). VideoNOC allows to analyze the impact of network conditions on video QoE, reveals video demand across the entire network, to develop and build better networks and streaming services. Despite, the platform disregard transport-layer and relevant RAN KPIs data and QoE inference on encrypted video traffic. 
\\
\item In the same vein, an online Machine Learning (ML) named ViCrypt is introduced \cite{seufert2019stream}, to anticipate re-buffering events from encrypted video streaming traffic in real-time. This approach, after it subdivides the video streaming session into a series of time slots, that have the same length. It employs a fine-grained time slot length of $1$ second (for a proper tradeoff between precision and stalling delay detection), from which, the characteristics are extracted. Afterward, they are used as an input to the ML model to predict the stalling occurrence. It should be mentioned that the initial delay and length of stalling events can be also be obtained. As an extension to the later work, the authors have demonstrated in \cite{wassermann2019let} that ViCrypt can additionally predict the video resolution and average video bit-rate accurately. As an extension to the later work, the authors have demonstrated in \cite{wassermann2019let} that ViCrypt can additionally predict the video resolution and average video bit-rate accurately. Also, Vasilev \emph{et al. } \cite{vasilev2018predicting} opted to build an ML model to anticipates the rebuffering ratio based on the hidden and context information to enhance the precision of prediction through Logistic regression.
\\
\item Lin et al. \cite{lin2017machine} applied a supervised ML and support vector machine to anticipate users' QoE by considering the number of active users and channel conditions experienced by a user. They classify a session in two categories (i.e., with or without stall events) based on cell-related information collected at the start of a video session. Considering the starvation events, mobile users experience them more than adaptive streaming and static users. As well these last, are more accurate and convenient to predict their starvation event. Similarly, a multistage ML cognitive method is developed by Grazia et al. \cite{de2017qoe}. Although, this model combines unsupervised learning of video characteristics with a supervised classifier trained to extract the quality-rate features automatically. Their model is supposed to exceed the other offline video analysis approaches.
\\
\item Orsolic et al. \cite{orsolic2017machine} proposes YouQ, an android application to prognosticate The QoE (i.e., stalls, quality of playout and its variations) employing ML relying on objective metrics like throughput and packet sizes extracted from the stream of encrypted packets.
Though, the promising result, the majority of the features depends on TCP, meaning that, in regards to UDP, these techniques probably will fail.
\\
\item Similarly the authors \cite {lopez2018deep}, suggested a QoE detector based on extracted data from networks' packets employing a deep learning model. The model is based on a combination of an RNN, Convolutional Neural Network, and Gaussian Process (GP) classifier.This classifier can recognize video abnormalities (i.g., black pixel, ghost, blockness, columns, chrominance, color bleeding, and blur) at the current time interval (in 1-second) and predicts them. The model is supposed to predict video QoE in a real-time environment; however, it could encounter a few issues like having a small amount of training data. 
\\
\item ECT-QoE framework \cite{duanmu2018quality} predicts at the instant the QoE of streaming over DASH, based on the expectation-confirmation theory and the video database, they have built. The model is presumed to defeat several models, especially when combined with the SSIMplus model. Despite that, ECT-QoE can be applied only to videos consisting of view segments. \\
\item Wu's model \cite{wu2018toward}, contrary to other propositions, examines the global intensity and local texture metrics extracted from a decoded video, to predict stalls event and assess the user's quality. The algorithm maps the normalized number and duration of stalls using linear combinations. When compared to other models (e.g., \cite{hossfeld2013internet}, \cite{mok2011measuring}, \cite{rodriguez2012quality}), Wu's proposition appears more consistent concerning subjective perception. \\
\item A cost-constrained video quality satisfaction (CVQS) framework is proposed \cite{li2018cost} to predict the quality expected, considering some metrics such as the high cost of data. Despite that, it indicates satisfactory results the accuracy of the CVQS could be impacted by the video encoder as well in their test the client can only obtain the next video segment after two seconds. 
 \end{itemize} 
There are a large number of standards, that offer indications on good and accustomed practices, for certain test applications, standards do not provide the best or most advanced method available, but it gives solid, common basis which is accessible to all, like ITU - International Telecommunication Union \cite{Streijl2014mos} (Table II). Furthermore, a survey \cite{pal2018survey} summarized various ITU- measurement methods to evaluate video streaming quality. 
\begin{table}[t]
\begin{center}
\caption{ITU Recommendations on Subjective and Objective measurement}
\setlength\tabcolsep{3.75pt}
\begin{tabular}{|c|c|c|c|}
\hline
& \textbf{\normalsize{Speech}} & \textbf{\normalsize{Audio}} & \textbf{\normalsize{Video}} \\
\hline 
\textbf{\rotatebox[origin=c]{90}{\normalsize{Subjective}}} & \thead{
P.85(2013)\cite{p85}\\
P.800(2016)\cite{p800}\\
P.805(2007)\cite{p805}\\
P.806(2014)\cite{p806}\\
P.808(2018)\cite{p806}\\
P.830(1996)\cite{p830}\\
P.835(2003)\cite{p835}\\
P.1301(2017)\cite{p1301}\\
P.1302(2014)\cite{p1301}\\
P.1311(2014)\cite{p1311}}
& \thead{
P.800(2016)\cite{p800}\\ 
P.830(1996)\cite{p830}\\
P.831(1998)\cite{p831}\\
P.911(1998)\cite{p911}\\
P.913(2016)\cite{p913}\\
P.1301(2017\cite{p1301})\\
P.1302(2014)\cite{p1302}\\ 
BS.1116-1(1997)\cite{BS.1116-1}\\
BS.1284(2003)\cite{BS.1284}\\
BS.1285(1997)\cite{BS.1284}\\ 
BS.1534(2014)\cite{BS.1534}\\
BS.1679(2004)\cite{BS.1679}}
& \thead{
P.910(2009)\cite{ITU-TRec.P.910}\\
P.912(2016)\cite{p912}\\
P.913(2016)\cite{p913}\\ 
P.917(2019)\cite{p917}\\ 
BT.500-10(2000)\cite{BT.500-10}\\
BT.1663(2003)\cite{BT.1663}\\
BT.1788(2007)\cite{BT.1788}\\
BT.2021(2012)\cite{BT.2021}\\
J.140(1998)\cite{J.140}\\
J.245(2008)\cite{J.245}\\
J.247(2008)\cite{J.247}}\\
\hline 
\textbf{\rotatebox[origin=c]{90}{\normalsize{Objective}}} & \thead{
P.563(2004)\cite{p563}\\
P.561(2002)\cite{p561}\\
P.562(2004)\cite{p561}\\
P.564(2007)\cite{p564}\\
P.862(2001)\cite{P.862}\\
P.863(2018)\cite{P.863}\\
G.107(2014)\cite{G.107}} &
\thead{
P.561(2002)\cite{p561}\\ 
P.564(2007)\cite{p564} \\
P.862(2001)\cite{P.862}\\ 
P.863(2018)\cite{P.863}\\
P.920(2000)\cite{p920}\\
P.1201(2012)\cite{P.1201}\\
P.1305(2016)\cite{P.1305}\\
P.1310(2017)\cite{P.1310}\\
P.1311(2014)\cite{p1311}\\
P.1312(2014)\cite{P.1312}\\
BS.1387(2000) \cite{BS.1387}\\
G.1070(2018)\cite{g1070}\\
G.1091(2014)\cite{G.1091}} & 
\thead{
J.343-rev(2018)\cite{j343}\\ 
P.1201(2012)\cite{P.1201}\\
P.1202(2012)\cite{P.1202}\\
P.1203(2017)\cite{P.1203}\\
P.1401(2012)\cite{P.1401}\\
BT.1683(2004)\cite{BT.1683}\\
BT.1866(2010)\cite{BT.1866}\\
BT.1867(2010)\cite{BT.1867}\\
BT.1885(2011)\cite{BT.1885}\\
BT.1907(2012)\cite{BT.1907}\\
BT.1908(2012)\cite{bt1908}\\
J.143(2000)\cite{j143}\\
J.144(2004)\cite{j144}\\
J.246(2008)\cite{j246}\\
J.247(2008)\cite{j247}\\
J.249(2010)\cite{j249}\\
J.341(2016)\cite{j341}\\
J.342(2011)\cite{j342}\\
G.1022(2016)\cite{g1022}\\
G.1070(2018)\cite{g1070}\\
G.1071(2016)\cite{g1071}\\
}\\ 
\hline 
\end{tabular}
\end{center}
\end{table}
\subsection{Hybrid Assessment}
\vspace{0.2cm}
According to \cite{brooks2010user},  QoE of a user's performance can be estimated based on objective and subjective psychological measures while using a service or product. Moreover, another approach exists that consists of a combination of subjective and objective assessment, referred to as The Hybrid approach. 
Using ML algorithms
\cite{xie2018survey, binsahaq2019survey}, statistics, and other fields. It could be employed in real time, and it is categorized as the most accurate approach since it decreases the weaknesses of previous approaches \cite{yang2018survey}.
\\ 

For instance, the Pseudo Subjective Quality Assessment (PSQA) was created to give similar results as perceived by human in real-time, as it provides an accurate QoE measurement \cite{piamrat2009quality, ghareeb2010hybrid}. PSQA is based on training a particular type of statistical learning approach, Random Neural Network (RNN). To evaluate the quality of the video, the IFs on the quality are selected to be used to generate several distorted video samples. Afterward, these samples are subjectively assessed. Then the results of the observations are employed to train the RNN in order to apprehend the relation between the factors that cause the distortion and the perceived quality by real humans. The training method is performed once, after that the trained network can be used in real time. A comparison study in \cite{piamrat2009quality} proved that PSQA is more effective than subjective (MOS), objective (PSNR), in the matter of time-consuming, manpower moreover it runs in real-time. Likewise, a further investigation was done \cite{ghareeb2010hybrid} in the context of Multiple Description Coding (MDC) video streaming over multiple overlay paths in video distribution networks, confirms the same result as in \cite{piamrat2009quality}. Because, after training MDC-compatible version of PSQA; PSNR could not evaluate, and its results did not change a lot corresponding to the Group of Pictures (GOP) size. On the contrary, PSQA module considered the size of GOP and differentiated if MDC is used or not. 
Nevertheless, this approach is not applied in wireless mesh networks. 
Fortunately, another tool called Hybrid Quality of Experience (HyQoE) can predict for real-time video streaming applications \cite{riker2011hybrid}. It takes into account six parameters percents losses in I frame, P frame and B frame, general loss, complexity, and motion. Comparing HyQoE to other tools, they have demonstrated that, PSNR algorithm does not take into consideration the human visual system and the MPEG structure during the assessment process. Also SSIM is inadequate to reflect the user opinion when different patterns of loss, motion, and complexity are analyzed, and that video quality mode generates low scores. 
\begin{equation}
SSIM(i, k) = \frac{(2 \mu_{i} \mu_{j} + c_{1})(2 \sigma_{ij} + c_{2})}{(\mu^{2}_{i} + \mu^{2}_{j} + c_{1})(\sigma^{2}_{i} + \sigma^{2}_{j} + c_{2})}
\label{SSIM}
\end{equation}
Where $\mu_{i}$ and $\mu_{j}$ are respectively, the average value in the block of the original and the distorted image. $c_{1}$ and $c_{2}$ are the variables that stabilize the division with weak denominator.
$\sigma^{2}_{i}$ and $\sigma^{2}_{j}$ are respectively, the variance in the block of the original and the distorted image. $\sigma_{ij}$ denotes the covariance of the block of the original and the distorted image. 
\\

\noindent HyQoE gives results quite similar to the one given by MOS. They believe that it can be used to optimize the QoE by improving the usage of the network's resources. 
 Likewise, Chen\emph{et al. }\cite{chen2009oneclick} proposed a framework that seizes the users' perception while using network applications named Oneclick. If ever the user is displeased, he can click a button to indicate his feedback. Then the collected data is analyzed to determine the user's perception under variable network conditions. The tool is supposed to be intuitive, lightweight, time-aware, and it is convenient for multi-modal QoE assessment and management studies considering its application independent nature. The framework considered to give the same result as MOS but faster. 
 %
 Furthermore, the authors in \cite{abar2017machine} employed four ML algorithm (i.e., Decision Tree, neural network, kNN, and random forest) to evaluate MOS value, Based on VQM and SSIM values (i.e., the effect of video distortion and structural similarity). Thus, to assess the performance of these algorithms, the Pearson correlation coefficient and the Root Mean Square Error are employed. According to the results, the Random Forest algorithm was the best in anticipating user perception. However, network parameters like transmission delay and response time are not taken into account.
 \\ 
 
\noindent MLQoE is a modular user-centric algorithm developed by Charonyktakis \emph{et al. } \cite{charonyktakis2015user}, based on supervised learning to correlate the QoE and network parameters such as average delay, packet loss, average jitter. The framework uses multiple ML algorithms (i.e., Artificial Neural Networks, Support Vector Regression Machines, Decision Trees, and Gaussian Naive Bayes.). The one that outperforms the others, as well as its parameters, will be selected automatically considering the dataset employed as input. According to their result, MLQoE can predict precisely the score of the QoE compared to other existing ML model.
As well, in \cite{letaifa2017adaptive} the authors have suggested a trained ML model that predicts the MoS value in SDN, based on network parameters (e.g., bandwidth, jitter, and delay), Their proposal seems to be efficient.
 %
\\

\noindent YoMoApp (YouTube Monitoring App) \cite{seufert2017unsupervised} is an under improvements tool. It monitors the application and the network layer (i.e., the total amount of uploaded and downloaded data, is logged periodically) for both mobile and WiFi networks streaming parameters. As well to obtain subjective QoE ratings from end-users (MOS). 
The data is, anonymously uploaded, to an external database. Then a map is generated from the uploaded data of all users to reveal how every network operator function and how to be employed to benchmark them. YoMoApp performs accurate measurements on an adequately small time scale (~1 second). They recommended that QoE measurements have to consider more extended video clips. However, the tool uses JavaScript, which can occasionally cause inconsistencies and errors. 
The latter was employed as well as another Android-based passive monitoring tool to investigate the precision of different approaches. Consequently, streaming parameters revealed high correlations to the subjectively than for the objective experienced quality, which proves that it is better suited for QoE monitoring.\cite{seufert2016application}. 
%
Also, authors in \cite {wassermann2019machine} have used YoMoApp to monitor video sessions and obtain several features from end-user smartphones (e.g., the signal strength and the number of incoming and outgoing bytes). They, using ML, introduce a lightweight approach to predict Video streaming QoE metrics such as initial delay, number, and the ratio of stalling and user engagement. According to their evaluation, network layer features is enough to get accurate results. 
Recently, \cite{bampis2017learning} propose an ML model called Video Assessment of Temporal Artifacts and Stalls (ATLAS). It uses an objective video quality assessment (VQA) method by combine QoE-related features and memory features sources of information to predict QoE. They have also adopted, a subjective assessment, LIVE-Netflix Video QoE Database\cite{bampis2017study} to evaluate their model. Although the model is only apt to deliver overall QoE scores and cannot be used for real-time bit-rate decisions. \\
\noindent To sum up, the hybrid approach can collect metrics simultaneously from both the network and user-end. Such methods would help to correlate the QoS metrics on the QoE and generate a better MOS prediction tool. 
Also, hybrid studies will allow the study of the impact of the variations in the performance of the network on the users' QoE \cite{juluri2015measurement}. 
\\
Moreover, little research has been conducted in this area. Like in \cite{bartolec2019network}, authors have examined the effect of user behavior (e.g., seeking, pausing, and video skipping) on the accuracy of the trained QoE/KPI estimation models. They have concluded that when including user's various interactions, much better results will be obtained. However, more studies should be done.
\\\\
In Table III, we have summarized a few measurement approaches (i.e., subjective, objective, and hybrid). We outlined the methods, techniques, and challenges for each one of them.
\begin{center}
\begin{table*}
\caption{QoE Measurement Approaches.}
\centering
\setlength\tabcolsep{5pt}

\begin{tabular}
{|>{\centering}m{0.9cm}|>{\centering}m{1.1cm}|>{\centering}m{1.1cm}|>{\centering}m{0.9cm}|>{\centering}m{1.3cm}|>{\centering}m{4.7cm}|>{\centering}m{1.2cm}|c|c|c|}

\hline 
\multirow{2}{1.2cm}{{Related literature}} & \multicolumn{3}{c|}{{Measurement Approaches}} & {\ \footnotesize Monitoring Point} & {Methods and Techniques} & {Complexity} & {Accuracy} & {Challenges} 
\\ \cline{2-4} 
& {Subjective} & {Objective} & {Hybrid} &  &  &  &  & \\
\hline\hline 
\cite{xu2014flow} &  & \checkmark &  & Network &\begin{minipage}[c]{4.7cm}\normalsize {Both CBR and VBR streaming were considered under static and fast fading channels. Ordinary differential equations were constructed over a Markov process, to determine the prefetching delay distribution and the starvation probability.}\end{minipage} & High & Yes & \begin{minipage}[c]{2cm}\small Computational complexity\end{minipage} \\
\hline 
\cite{hobfeld2014survey} & \checkmark &  &  & User & \begin{minipage}[c]{4.7cm}\normalsize {Platforms and web-based frameworks performed online gather submitted opinions about the subjected test from different participants around the world.} \end{minipage} & Low & Yes & \begin{minipage}[c]{2cm} \small users' fairness\end{minipage}\\
\hline 
\cite{chen2009oneclick} & &  & \checkmark & User &\begin{minipage}[c]{4.7cm}\normalsize {Collect data after the users' feedback, then analyze them to determine the users' perception. }\end{minipage} & Low & Yes & \begin{minipage}[c]{2cm} \small users' fairness\end{minipage}\\ 
\hline 
\cite{xu2014analytical} &  & \checkmark &  & Network & \begin{minipage}[c]{4.7cm}\normalsize {They work on a slow channel fading or shared by multiple flows over a wireless network, modeled by a continuous time Markov process. After formulating the ordinary differential equations they solve them with Laplace Transform, to be presented as starvation probability and the mean continuous playback time. }\end{minipage} & High & Yes & \begin{minipage}[c]{2cm} \small Uncontrollable network conditions \\\\ Computational complexity\end{minipage} \\
\hline 
\cite{staehle2010yomo} & & \checkmark & & User & \begin{minipage}[c]{4.7cm}\normalsize {Detect YouTube's flow and analyze the packets in order to calculate the buffer status, then monitor it constantly; if the status falls below a critical threshold, it raises the alarm. }\end{minipage} & High & Yes & \begin{minipage}[c]{2cm} \small User's fairness \\\\
 User's privacy\end{minipage}\\ 
\hline 
\cite{juluri2011pytomo} &  & \checkmark &  & User &\begin{minipage}[c]{4.7cm}\normalsize {Computes and stores the downloaded statistics in a database of each video, by resolving the IP address of the video server; that is used to perform the analysis. }\end{minipage} & High & Yes &\begin{minipage}[c]{2cm} \small Data limitations\end{minipage}\\ 
\hline 
\cite{eckert2013qoe} &  & \checkmark &  &Network &\begin{minipage}[c]{4.7cm}\normalsize {First, they extract the playout timestamp, and then the algorithm calculates the actual buffer fill level and the duration of the stalling event. }\end{minipage} & High & Yes &\begin{minipage}[c]{2cm} \small Access to the user's device \\\\ Algorithm limited to YouTube
 \end{minipage}\\
\hline 
\cite{liu2014software} &  &  & \checkmark & User and Network & \begin{minipage}[c]{4.7cm}\normalsize {After collecting the viewers' experience, through a mathematical model, QoE scores are evaluated. }\end{minipage} & High & Yes & \begin{minipage}[c]{2cm}\small Computational cost \\\\ User's fairness\end{minipage}\\ 
\hline 
\cite{riker2011hybrid} &  & & \checkmark & User &\begin{minipage}[c]{4.7cm}\normalsize { HyQoE evaluates the quality of the video, based on static and trained learning (random neural network) over a wireless mesh. }\end{minipage} & High & Yes & \begin{minipage}[c]{2cm}\small Computational complexity\end{minipage}\\
\hline 
\end{tabular} 
\end{table*}
\end{center}
\section{Controlling Quality of experience }
\vspace{0.2cm}
As previously presented, various metrics influence the QoE. In this section, several approaches and observations will be discussed to enhance and control the QoE of video streaming services. Some may presume that increasing QoS, means precisely a higher QoE as stated in \cite{matulin2014subjective}. Except that the user could be content if he is expectations and requirements are fulfilled, especially if the context of the video is interesting. The previous findings were confirmed by \cite{kara2016one} as their results indicated that even frame freezes and shorter playbacks are acceptable by viewers. \\

\noindent Although it was proven that as the number of starvation increase the experience decrease, which the user is unable to endure and finally deserts the video \cite{chihani2014user}. Hence, to avoid starvation, prefetching/Start-up delay and re-buffering delay, a model was proposed in \cite{ye2014computing}, to optimize the QoE by computing the optimum start-up threshold that influences the number of starvation, which allows the content provider to achieve its QoE requirements choosing the right QoE metrics and to avoid starvation. Likewise, authors in \cite{xu2014analysis}, when analyzing the buffer starvation, have suggested that service providers should configure different start-up threshold for different categories of media files. Furthermore, based on the observations in \cite{xu2016modeling} they advice network operators, that to enhance the QoE of short views, they should be configured in a higher scheduling priority to reduce the starvation significantly and start-up delays, in the other hand the probability of starvation will slightly increase for long views. However content providers are unwilling to share statistics of views with network providers. \\

 \noindent The authors in \cite{bouraqia2018solving} adopted Lagrange Multiplier, after studying 
 the probability of starvation ($p_{s}$) of different file distribution to exploit the trade-off 
 between $p_{s}$ and the start-up delay. They were able to optimize the start-up delay by 40\%. In contrast, dynamic adaptive bit-rate was not considered in their scenario. In the same manner, 
 another work \cite{kim2018kkt} used KKT-conditions based on a Resource Allocation Algorithm 
 \cite{bazaraa2013nonlinear} to optimize the problem (i.e., reduce the occurrence of stalling 
 events, assure fairness among users (whether utilizing dynamic adaptive streaming or not)). 
 Compared to other proposals (e.g., Proportional Fair Resource 
 Allocation \cite{kwan2009proportional} and Base Station Optimization\cite{xiao2016qoe}) theirs 
 indicate better performance. 
For example, in a disturbed traffic network, authors \cite{laghari2018measuring} proposed to keep the packet reordering percentage below $20\%$ to maintain an acceptable level of QoE. Still, they have streamed the video using UDP protocol in their study. 
\\

\noindent The streaming service has adopted a new protocol that answers to the massive demand on network requirements like bandwidth, entitled DASH\cite{timmerer2017quality, stohr2017sweet}. It is proved to adapt the quality of the requested video, based on the current bandwidth and devices qualification, but it is affected by many factors based on \cite{liu2013user, barman2019qoe}, initial delay, stalling and level variation (frame rate, bit-rate and resolution), besides other factors like video length and the number of motions in the video. 
Consequently, to derive an effective the trade-off between the network variations and dynamic videos streaming behavior, they \cite{burger2018generic} introduce a queue-based model to analyze the video buffer (GI/GI/1 queue) with pq-policy (pausing or continuing the video download) using discrete-time analysis. Suggesting to adjust the buffering thresholds according to the bandwidth fluctuations to reduce the stalling vents. In the same aspect, authors \cite{schwarzmann2018evaluation}, after studying the impact of variable and fixed segment duration (HAS streaming services commonly use segments of equal duration) on the stalling probability, proposed a variable segmentation approach that effectively increases the content encoding (i.e., reduced bit-rate per video clip. However, the segment duration can affect the QoE of the streaming behavior of DASH. 
Besides, authors in \cite{li2018cost} suggested a trade-off between profit and service, to network operators and mobile providers. It states that based on several metrics like cost of data and encoding, they can decide the suitable quality level to transfer data to the end-user and thereby, reduce the video storage and optimize resource allocation. 
In the same context, the framework named QUVE \cite{kimura2017quve} intended to increase the QoE of video streaming services. It comprises two principal sections the first approach, the QoE estimation model, considers encoding parameters, re-buffering conditions and content time to assess the QoE for Constant Bit Rate (CBR) video streaming. The second, QoE parameter estimation approach, it predicts the network quality, re-buffering time and count for the proposed model. The results attest that QUVE is adequate to improve the QoE by choosing the adequate encoding based on a user network conditions. 
 \\
 
\noindent In another context, users usually find it troubling to decide the next segments quality level to maintain a high QoE. Thus an extension of DASH player is presented \cite{bokani2016implementation} to make a decision based on Markov Decision Process (MDP) called MDP-based DASH. It requires a bandwidth model and a learning process, so after adequate training, the player parameters are tuned to be employed. It is shown that adopting MDP to adapt video quality will reduce notably the video freezing and buffering events. 
\\

\noindent There exist also a bit-rate switching mechanism permitting users to choose among different switching algorithms to control the starvation probability, which is difficult to define its behavior, as the wrong choice affect the QoE. In \cite{xu2014analytical} a framework is proposed to assist the user in finding the optimal bit-rate to optimize the QoE, taking into consideration all the future occurrences. 
Also to provide the QoE expected from video streaming HTTP Adaptive Bitrate (ABR) was adopted, caching many streaming files to meet up with the QoE requirements. ABR encountered a problem of storage to control it, an optimal subset of playback rates that would be cashed is chosen. As a solution to this problem, the authors in \cite{liu2014software} developed a model for QoE driven cache management to offer the best QoE and avoid the content storage to be filled up rapidly. 
\\

\noindent Regarding the increase in energy consumption in a cellular network and mobile devices authors in \cite{zhang2016profiling}, have conducted a study on the subject. They have asserted that to maintain a good balance between QoE and energy consumption, while watching a video from a mobile phone over Long-Term Evolution (LTE) networks, a new design of video streaming service will decrease the energy consumption by $30\% $. Though; some points (increasing the length of video segments, increasing the buffer size, the strength of the signal and using appropriate DASH sittings) should be taking into consideration. 
In another paper by Song \emph{et al.} \cite{song2016edash}, they propose an Energy-aware DASH (EDASH) framework over LTE to optimize network throughput and to find an excellent balance between the energy consumption of the users' device and the QoE, that proves based on their experiments, its efficiency. 
The authors in \cite{colonnese2016timely} have determined the mathematical formula expressed by two QoE metrics (video rate, the probability of timely delivery of video packets), in order to compute the probability of time delivery of DASH over a wireless access cell (LTE) to determine the bandwidth assigned to the mobile user to maintain a satisfactory QoE. 
Moving cell phones between wireless access networks make it hard to maintain a good QoE. Thus in \cite{park2016quality} they have proposed an adaptive streaming protocol consisting of network adaption and buffer management block that dynamically adapts the bit-rate according to network conditions fluctuations, to provide a stable QoE over 5G. The protocol is designed independently of the operating system (OS) version and CPU performance of the mobile device. The result indicates that the proposed protocol seemed to enhance the users' QoE, as it has been deployed commercially in South Korea for more than five years over commercial LTE/3G and wifi networks. 
\\

\noindent To improve DASH efficiency under different network conditions, they suggest \cite{zhao2017study} a dynamic adaptive algorithm that can be utilized in both bandwidth and buffer based methods. It depends on current bandwidth fluctuation to choose the best quality video, guarantees the continuity and real-time video streaming to keep a high QoE. To test their model they have to utilize Google ExoPlayer \cite{Exoplayer} an Android-based mobile DASH as a video player for Android. The results obtained attest that the approach attains a significant average QoE and performs steadily under various networks as no rebuffering happens except in the initial buffering stage ($~0.35$ seconds ). 
\\

\noindent In addition, to address the problem of network delays for CBR and Variable bit-rate (VBR) over 5G mobile networks. In this paper \cite{mushtaq2016qoe} they describe an analytic method that addresses this challenge. Also, the authors present a method to compute the users' QoE based on an exponential hypothesis for streaming traffic using delay and packet loss rate as metrics. This approach decreases the network delays of traffic by less than 1 ms, therefore improve the QoE. 
\\

\noindent Furthermore, in some bidirectional streaming services, the up-link capacity might also be required as much as downlink capacity. For instance, the authors of \cite{nunome2012qoe} propose a piggyback mechanism for audio-video IP transmission over the uplink channel to enhance the QoE, which seems to perform well. The result obtained shows that the mechanism is rather more effective in adaptive allocation schemes than under static allocation schemes. However, it seems Nunome and Tasaka have tested their proposition on other classes of contents. 
\\

\noindent Dutta \emph{et al. } \cite{dutta2016qoe}, to face the challenges encountered in 5G networks (i.e.,  arranging the connectivity of high data rate to an expanding mobile data traffic), suggest an approach to allow the cloud infrastructure to dynamically and automatically change the resources of a virtual environment, to use the resources efficiently and to provide an adequate QoE. The approach seems to be able to ensure a real elastic infrastructure and promising in handling unexpected load surges while reducing resource, demanding real-time values of PSQA. 
\\

\noindent Other research efforts suggest that a better quality perception might be met when the quality should be controlled. In \cite{zinner2010controlled, fiedler2010quality}, the authors apply provisioning-delivery hysteresis for QoE in video streaming case, in order to predict the behavior of the throughput and the QoE to control the quality, using the SSIM. Another mechanism \cite{taleb2012qos} is proposed to control the quality, as congestion degradation affects QoS which impacts thereby the QoE of users. The authors introduce an Admission Control (AC) mechanism based on QoS and QoE metrics, using a joint QoS/QoE that is predicted by a QoS/QoE mapper. Based on these metrics the AC decides whether the user should be accepted within the small-cell network on not. Though the results obtained are encouraging, AC is only simulated and has not been implemented in realistic network as far as we know. 
In addition, an introduction of SELFNET 5G project \cite{nightingale2016qoe} provides a self-organized capability into 5G networks achieving autonomic management of network infrastructure. It designs and implements an adaptable network management framework to provide scalability, extensibility and smart network management reducing and detecting some of the network problems. The framework improves the QoE also and reduces the operational expenditure (OPEX). 
\section{Bringing QoE at the Edge}
\vspace{0.2cm}
In a typical scenario, when a mobile device requests a video content, it is issued from the servers of CDN, then crossing the mobile carrier Core Network (CN) and Radio Access Network (RAN). Clearly, a massive number of simultaneous streams would generate a colossal demand at backhaul side. Moreover, the wireless channel uncontrollable conditions (e.g., fading, multi-user interface, peak traffic loads, etc.) might be a challenging issue for the monitoring of user's QoE and would be an additional load on the cellular network. Yet, delivering a streaming content is rather difficult, giving that the channel between servers providing the desired content and users can cause delays when transporting data, which would impact the user's experience. Bringing the content closer to the user via caching promises to overcome several obstacles like the network load and delays resulting in an enhanced QoE \cite{shefkiu2017quality}.
\\

\noindent To improve users' QoE when using dynamic rate adaptation control over information-centric networks, StreamCache \cite{li2016streamcache} is proposed. This latter periodically collects statistics (i.e.,  video requests) from edge routers to make a video cache decision. The results indicate that this approach offers a near-optimal solution for real-time caching as it enhances the QoE by increasing the average throughput. However, the cache size at routers level might influence the performance. Also, a Mobile Edge Computing (MEC) scheme was suggested \cite{ge2016qoe} to permit network edge assisted video adaption based on DASH. The MEC server locally caches the most popular segments at an appropriate quality based on collected data from the network edge (i.e.,  throughput, latency, error rate, etc.). To solve the problem of cache storage a context-aware cache replacement algorithm, replaces old segments by new popular ones, which leads to maximizing the users' QoE as it ensures a steady playback minimizing frequent switching.
Proactive service replication is a promising technique to decrease the handover time and to meet the desired QoE between different edge nodes. However, the distribution of replicas inflates resource consumption of constrained edge nodes and deployment costs. In \cite{farris2017optimizing}, the authors have proposed two integer linear problem optimization schemes. The first scheme aims to reduce the QoE collapse during the handover; whereas the second scheme  aims to reduce the cost of service replication. Evaluating this scheme in MEC, mentioned above, the authors believe the effectiveness of their solutions as well they could provide more information about the network (i.e.,  predict the user's mobility pattern).\\ 

\noindent Furthermore, to manage calls' handover 
in wireless mesh networks, a testbed technique 
\cite{brito2017improving}, combines RSSI (measure the strength of the received signal) and RTR (as 
an indicator of transmission rate quality) to 
compute the quality of a wireless link (every 1 
second). This procedure allows monitoring and 
takes decision of handover (select the access 
point with the highest quality level). On one 
hand, this scheme is assumed to improve the QoE 
by 70~\%. On the other hand, it might increase 
the amount of updates, delays besides it 
disregard variable bit-rate (VBR). \\

\begin{figure}[!ht]
\centering
 \includegraphics[scale=0.35]{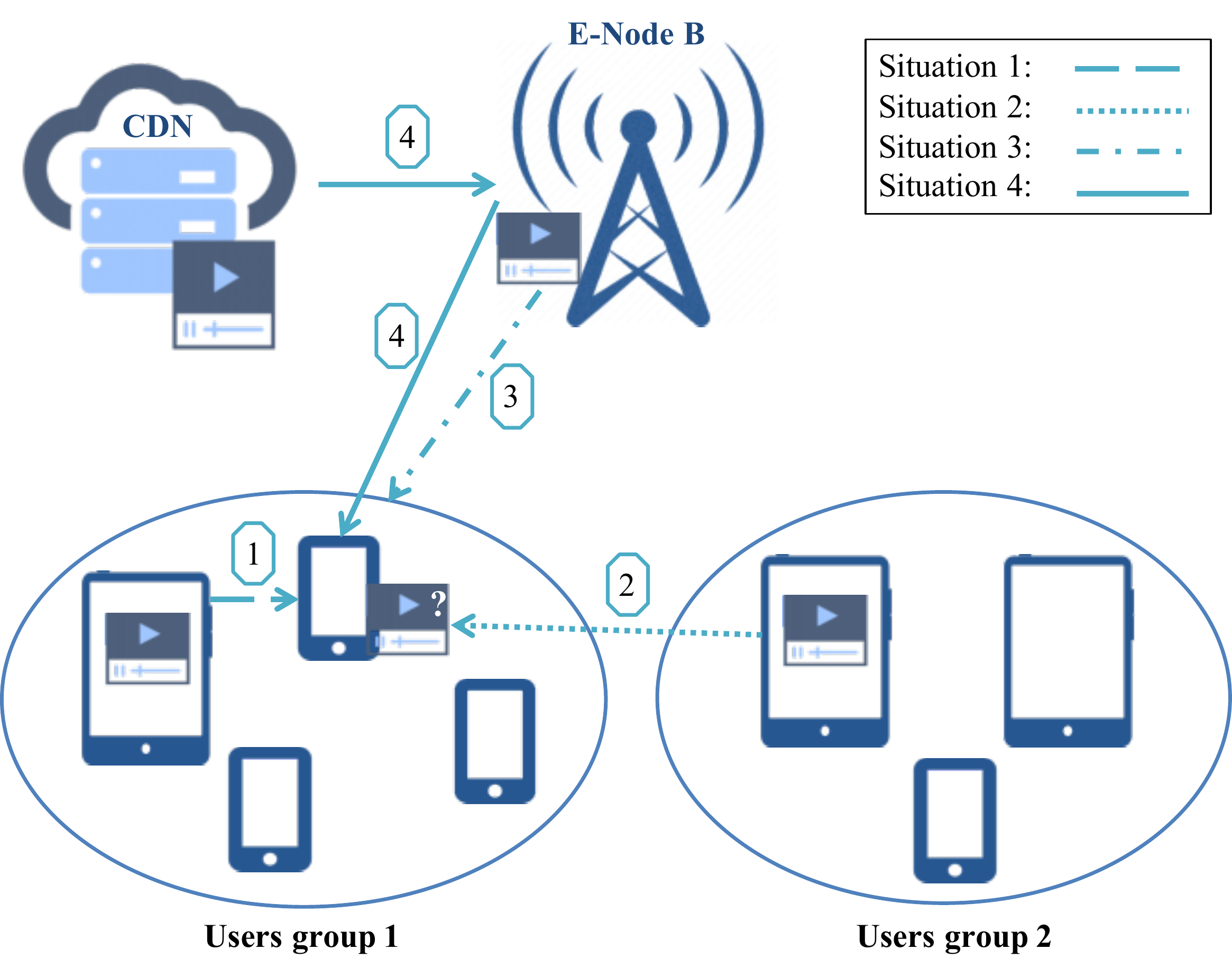} 
\caption{Request a video file from a neighbor device and get served.}
\label{Caching}
\end{figure}

 \noindent In another research piece \cite{timmerer2016transcoding}, the authors propose a cloud encoding service and a Hypertext Markup Language revision 5 (HTML5) for adaptive streaming player. The built player is a client framework that could be integrated into any browser. The server side is implemented within a public cloud infrastructure. It has been claimed that this scheme promises the elasticity and the scalability needed to suit the clients, although this approach is specifically destined for MPEG-DASH. 
 Due to a rapid growth of mobile data traffic (e.g., mobile videos), the authors of \cite{molisch2014caching} develop some optimum storage schemes and some dynamic streaming policies to optimize the video quality, combing caching on a device and D2D communication to offload the traffic from cellular network as well as the available storage on mobile devices. They introduce a framework called reactive Mobile device Caching (rMDC). Hence, instead of requesting a video from the base station, in D2D caching network, the user can request it from neighboring users and might be served over an unlicensed band.
In such a way, D2D candidates are detected before starting communication sessions between devices, using assigned beacon (synchronization or reference signal sequence) resources by the network. This beacon will be broadcast in the cell area to allow devices to advertise their presence and identify each others \cite{6163598}. Thereby, in the occurrence of a video request, the device starts searching its cache and afterwards, it explores the neighbors' caches locally to retrieve the desired video. If it does not appear, the cache agent at the e-NodeB attempts to locate another mobile device in another group that belongs to the same area. Finally, if the video is not located in any other neighboring device, the cache agent will program to get the missing chunks of the videos from the cache of the e-NodeB if they exist, else they will be downloaded from the CDN. Figure -\ref{Caching}- indicates the different transition that the mobile device might take before obtaining the desired video.
Here, the authors have proved that using rMDC along with user preference profile-based caching, their framework seems to perform well and reaches high network capacity and better video QoE for mobile devices. Besides, the distance between the mobile device and the server hosting the video might be long and could impact the QoE. In \cite{abd2015optimization}, the authors propose two mechanisms for files duplication: 1) caching (duplicate copy of a file in different places); and 2) fetching (retrieving the video to another place or zone) simulated separately in different scenarios. Based on the observed demand on a given file, it is selected and the duplication algorithm is activated to duplicate it at the operators sharing server, to be closer and more accessible to the user with good quality and minimum cost. The content fetching seems to be more efficient than caching, and combining these mechanisms might produce even better results. \\

\noindent To efficiently bring a given content to end-users with a satisfactory QoE level, the CDN administrator should ensure that this content is strategically stored/cached across the Web \cite{dutta2016fly}, \cite{frangoudis2016architecture}, as this profoundly impacts the user experience. Storage policy also influences the cost, both in terms of CAPEX and OPEX, to be paid by the CDN owner. It also plays a crucial role in offering of CDN as a service (CDNaaS) \cite{retal2017content}. CDNaaS is a platform that could establish virtual machines (VMs) over a network of data centers and provides a customized slice of CDN to end-users. Moreover, it can handle a significant number of videos through caches and streamers hosted at different VMs. The authors formulate two linear integer solutions for VM placement problem, that was implemented using Gurobi optimization tool, Efficient Cost Solution (ECS) and Efficient QoE Solution (EQS). In terms of maximizing QoE, EQS algorithm shows the best performance. However, regarding time, ECS algorithm exhibits better performance, disregarding the number of data centers and the number of flavors per location. \\

\noindent In order to deliver virtual server resources in a CDNaas architecture, \cite{koskimies2017qoe} presents a QoE estimation solution that can be employed as a part of a QoE-aware system. The developed system discovers how many users can simultaneously be handled by a server while granting a satisfactory service quality level. It aims to capture how the QoE of a video stream is affected by different factors. The results, based on PSQA, reveals that stream segment duration is an influential factor, and needs to be taken into account throughout resource optimization. The system might be used as a part of the QoE-optimized resource. However, the authors seem to have overlooked the effect of network bandwidth. \\ 

\noindent From a different perspective, an optimal rate allocation was designed by \cite{vo2018optimal} to limit the co-channel interference and manage resources between D2D and cellular users. They are using a joint encoding rate allocation and a description distribution optimization forwarded to BS and D2D users (predefined candidates, who already cached the content, and who are selected based on their available storage and battery level) before transmitting video segments to the requester. They believe that the scheme improves the QoE of video streaming delivery. Despite, the authors did not consider the additional delays that would be generated by the optimization process at the BS level. Also, a dynamic allocation method is adopted in \cite{he2018qoe}, implementing the shortest path tree, to allocate joint resources (i.e., video streaming, files, etc.). The results conclude that selecting the appropriate transmission rate and the dynamic allocation, could result in an enhanced QoE. Still, the authors assume the content chunks have the same size and the transmission rate is the same for all active nodes, which is not true in real networks scenarios. The end-to-end communications in Next-Generation Networks (NGN) between users and application servers may cross different networks belonging to different operators and implementing different technologies, which is challenging in terms of measuring, monitoring and managing the QoE. \\

\noindent According to \cite{zhang2011assuring}, optimizing the QoE requires that some factors should be considered like application-level QoS, allocated capacity, customer premise factors and subjective user factors. These factors are hard to figure out due to the difficulties of measuring subjective factors, and some of the elements degrading the QoE may not be available for diagnoses. Moreover, crossing several heterogeneous networks/links makes it hard to determine the element that induces a poor QoS level. In this regard, the authors build a framework that can be implemented in NGN, where the user is able to report the perceived QoE and QoS via software, which allows the operator to allocate the resources and reconfigure them accordingly. Nevertheless, the cost in terms of reporting, and the changing in the parameter might affect the performance significantly. Moreover, some networks might refuse to join and prefer to manage their QoE independently. A new dynamic and a reconfigurable Machine-to-Machine (M2M) network is proposed by \cite{yano2013dynamic} where the two metrics are introduced allowing to manage the wireless network, operational quality of applications and efficiency of wireless resource utilization. These metrics allow the network to cover more applications running with higher QoS level and enhanced QoE metrics. The authors consider a multiple layer sensing to the proposed system, so as the platform collects information from each wireless node in the wide area and then forwards the resulting control information to the management network entity. Thereby, the network management decides to optimize the network topology and so on.\\ 

\noindent Mobile network operators have a limited spectrum/bandwidth, and they pay billions of Dollars to obtain time-limited licenses. Hence, obtaining efficient spectrum usage to get the required capacity is of great interest both for operators and end-users. Thus, communication network needs to increase the capacity to cope with the growing demand for data transmission. The authors of \cite{liu2014device}, have described and clearly formulated this problem, and the new areas of research on infrastructureless communication (e.g., D2D, M2M, etc.) and small-cells. They also emphasize some innovative spectrum management options, that permit more flexible use of spectrum while enabling D2D communication and deploying small-cells to be candidates to ease such a flexible usage of spectrum. \\

\noindent Long Term Evolution-Advanced (LTE-A) significantly enhanced the spectral efficiency. Yet, the imbalanced traffic distribution among different cells and the severe congestion of some of them still a challenging issue. Techniques like smart-cells \cite{nakamura2013trends} and biasing \cite{bou2010enhancing} seem to be promising and might partially solve such a problem. Yet, although they cannot deal with real-time traffic distribution, authors of \cite{liu2014device} propose a D2D communication-based load balancing algorithm to increase the ratio of user equipment (UE) that can access Internet at the same time. This mainly helps offloading traffic of macro cells via small cells. However, unfortunately, this algorithm could only be utilized for network applications/services, and is not adapted to streaming service as it suffers from some drawbacks like security issues and interference management. \cite{chen2017qoe} presents a resource management algorithm Media Service Ressource Allocation (MSRA). This scheme schedules limited cellular network resources based on content popularity, while considering channel conditions and packet loss rate of D2D direct links. It also allows to achieve an interesting tradeoff between the amount of video service delivered and available cellular resources. Compared to other schemes, MSRA benefits from a rapid users' services distribution adjustment, reduces the impact of D2D underlying interference and enhances the QoE level. For better QoE fairness over services in LTE/LTE-A, a self-tuning algorithm \cite{oliver2016self} is proposed. The key idea is to repeatedly change the service priority parameters to (re)prioritize services, and guarantee that all users achieve the same average QoE regardless of the type of running service. Depending on whether the objective is to improve the average service QoE or the individual QoE, the authors present two algorithms: 1) QoE unweighted approach, and 2) QoE weighted approach. This way, the appropriate algorithm is selected according to the preferred objective function. Thus, if fairness between services is desirable despite the number of users per service, the unweighted algorithm is used. Otherwise, the weighted algorithm priorities the popular services to enhance the user's QoE by around 15\%. \\

\noindent LTE wireless network supports most of M2M Communication classes. Yet, it faces many challenges like dealing with a massive number of M2M devices without influencing the users' QoE. While LTE scheduler plays an important role, it does not distinguish between M2M terminals and legacy UEs. It follows that the radio resources scheduler which turns to be in favour of M2M terminals over user equipment. As a solution \cite{abdalla2013qoe} suggests an M2M-aware hybrid uplink scheduler to balance the radio resources allocation, which provides adequate scheduling of M2M terminals without affecting standard UEs and the perceived QoE. Machine Type Communication (MTC) allows communication of machines or devices to machines over mobile networks. It is expected to exceed billions of M2M connections, still it might overload the system when a massive number of MTC devices attempt to connect simultaneously to the mobile network. The problem is addressed in \cite{taleb2014lightweight} regarding a Lightweight Evolved Packet Core (LightEPC) to organize the on-demand creation of cloud-based lightweight mobile core networks dedicated for MTC and to simplify the network attach procedure, by creating an NFV MTC function that implements all the conventional procedures. The latter scheme is shown to exhibit some nice efficiency and scalability features.
\section{Open Issues}
\vspace{0.2cm}
Although QoE modeling has gained a tremendous attention recently, it is still a challenging topic due to its multidisciplinary and subjective nature. For instance, it is hard to get access to operators' network data and traces, which makes it hard to experiment in realistic environments. Also, lack of open source video database to test quality metrics is being a high barrier towards understanding, assessing, improving and controlling the QoE. 

\subsection{Need to develop robust and realistic models}

Most of existing QoE models consider only a few parameters and not all QoE impacting factors. Whilst many IFs have been identified, such as user and context (e.g., habits, cultural background, environment, etc.), should be taken into account to design a robust and holistic model. Moreover, most reviewed articles do not offer a full study on the complexity of the proposed models from resource allocation (e.g., computing capacity, storage, energy consumption, etc.) perspective, specifically for handsets like smartphones, reducing the performance of the suggested assessment applications.

\subsection{Need to consider powerful tools to predict and assess QoE}

Throughout this article, we have surveyed a long list of methods aiming to assess QoE and control it. Unfortunately, none of them seems to perform well under general realistic settings. Namely, most of schemes suggested in related literature are only valid for some specific cases, under some strong assumptions in terms of content, user profile, handset, environment, etc. Artificial intelligence and machine learning algorithms have been recently used to measure the QoE objectively or to improve it. For instance, DASH uses machine learning to set the appropriate resolution and/or bit-rate according to the channel state. Allowing this way to continuously track the QoE and proactively take appropriate actions to keep good user experience. However, unfortunately, few works have used machine learning for hybrid assessment which gives similar results to subjective measurement approach. This performance collapse is probably due to the massive amount of required data, computation, verification and the complexity of the training model. We believe this research direction is still in its infancy and needs to be explored in depth. Furthermore, other powerful tools could be used to provide a better understanding of the QoE evolution over space and time. For instance, we believe mean-field game theory is a promising framework that may allow to model and track the QoE variation, while capturing the interaction among active users. More precisely, mean-field game theory turns to be very efficient in analyzing the behavior of a massive number of actors under uncertainty (e.g., random channel, random number of active users, unknown locations of attractive contents, etc.), by averaging over degrees of freedom allowing hereby to deal with a much simpler problem equivalent to the original complex problem. 
\subsection{Need to consider human at the center of service design process}

Recently new applications/technologies have emerged \cite{lounis2018neural}, requiring an unprecedented  requirement in terms of high data rate and extremely low latency. Consequently, promising the best possible experience is non-trivial due to diverse factors. As future applications like Virtual Reality (VR) Augmented Reality (AR), Mission Critical (MC) services, Tactile Internet (TI) and teleportation will require a colossal amount of resources, end-users will keep asking for high QoE while using these apps \cite{akhtar2019multimedia}. The international telecommunication union\cite{vision2015framework} has highlighted numerous requirements for the developing agreement on the usage states and needs of the emerging services (e.g., e-health, remote tactile control, etc.). Additionally, the technical infrastructure developments of the 5G communication systems have been evaluated in the context of recent system requirements (e.g., high bandwidth, low latency, and high-resolution content) and new experiences of users such as 4K resolution video streaming, TI, and AR/VR. \\

\noindent AR allows people to add digital elements into their existing environment (e.g., Snapchat, Instagram, PokemonGO, etc.). Billions of mobile users already heavily used, and many companies like Apple and Google Glass, Microsoft HoloLens are encouraging developers to build AR-Apps. Conversely, VR changes the real world into a virtual one requiring specific special hardware such as Oculus Rift gear (expensive and not-portable), which is slowing down its adoption rate by end-users. Moreover, TI \cite{holland2019ieee} will combine many technologies such as mobile edge computing, AR/VR, automation, robotics, telepresence etc.. Also, it will permit the control of the Internet of Things (IoT) in real time while moving and within a particular communication range. Further, a new dimension will be added to human-to-machine interaction by enabling tactile and haptic ( sense of touch, in essence, the manipulation and perception of objects utilizing touch and proprioception) sensations, and at the same time transform the interaction of machines. Therefore, assessing the QoE of such an application would need to consider all new parameters and will extremely specific QoS (e.g., ultra-reliability and low-latency) \cite{kabalci20195g}. 
\\

\noindent Inevitably, these emerging applications are changing our daily life and surrounding environment (e.g., home, work, etc.), which impacts our perception and understanding of space and time. Indeed, numerous study such as \cite{cai2019tablet} have proven that AR increases the learning ability. Earlier to this, more research must be conducted in various demographic, geographic areas. To incentivize users to experience and interact with immersive environments, it is fundamental to provide seamless services with perfect audio/video data processing capabilities. The most crucial performance metrics of these applications are typically high energy consumption and long processing delay \cite{ren2019edge}. To overcome the computational resource shortage of mobile devices novel techniques like mobile cloud computing and mobile edge computing are to be examined to allow users offloading the intensive computation tasks to several robust cloud servers. However, for more efficiency, a convenient edge-to-cloud architecture should be constructed. In this aspect, machine learning techniques can be applied to approach these difficulties possibly by using available traces. For example, to anticipate computational requirements so that devices could minimize latency, proactive scheduling of the computational resources could be performed in advance \cite{zhang2019thirty}.\\
 
\noindent As mentioned earlier, TI, MC, VR and AR, are new classes of applications that completely change the way we interact with reality. It is essential to keep in mind, that they can massively impact the brain, and affect its perceptions and reasoning, directly in an obvious manner 
(e.g., motion sickness, addiction, discomfort, eyestrain, nausea, migraine, etc.) \cite{you2019fog}.
Thus more studies have to consider these critical issues.
 \subsection{Economics of QoE}
 
 Economics of telecom services has reached maturity as a tremendous research effort has been spent in developing joint QoS and pricing models. Most of these models capture the interaction among competing operators over a shared market under homogeneous services and inhomogeneous services. However, all these models only consider strategic pricing for delivered QoS, and only deals with optimizing CAPEX and OPEX.  Thus, interactive models considering QoE and its influencing parameters are still to be build. 
More precisely, charging end-users according to the QoE they receive is of great importance. Of course, the pricing is assumed to depend on the delivered QoS but also on the end-users' satisfaction level and context. A deep analysis of the interaction among content provider, service provider, network provider, broker and end-users is becoming of grand importance. This interesting research direction is highly inter-disciplinary as it involves: economics, logistics and demand-supply optimization, flow theory, cognitive science, psychological and behavioral science. 

\section{Conclusion}
\vspace{0.2cm}

In this article, we provide a comprehensive literature review on QoE, by presenting standard definitions as well the influencing factors of QoE, that depends mostly on the type of network, the type of device, content, services and users. Next, we list major tools and techniques allowing to monitor and measure/estimate the QoE of a given service. We also discuss the challenges encountered in wireless networks and mobile networks (e.g., LTE, LTE-A and 5G), such as network capacity and varying channel conditions. Then, we exhibit most impactful solutions from literature. Many improvement mechanisms and controlling approaches with promising potential and even effective, are also cited and analyzed. \\

\noindent With 5G being deployed around the world, providing responsive networks able to grant high throughput and low latency is not a challenging issue anymore. However, supporting extremely latency/reliability demanding applications such as VR/AR and tactile Internet is still to be addressed. Thus, we believe considerable research efforts need to deal with developing efficient mechanisms allowing to meet these requirements.


%





\ifCLASSOPTIONcaptionsoff
  \newpage
\fi



%


\bibliographystyle{ieeetr}
\bibliography{mybib}

%

%
\begin{IEEEbiographynophoto}{Khadija Bouraqia}
received the Master degree in Networks and
Telecommunications (2009) from University Chouaïb Doukkali (El Jadida,
Morocco). Currently, she is Fifth-year ph.d. Student at the national higher school of electricity and mechanics (Ensem) She served as a reviewer for international and national conferences (WF-IoT 2019, WINCOM’18, DocSI’18). Her current research interests include quality of experience, 4G and 5G networks, Streaming services, learning algorithms, and game theory. She has three conference publications and a journal under submission.
\end{IEEEbiographynophoto}

\begin{IEEEbiographynophoto}{Essaïd Sabir}
received the B.Sc. degree in Electrical engineering Electronics and
Automation (2004) from Mohammed V University (Rabat, Morocco) and the M.Sc in Telecommunications and Wireless Engineering (2007) from National Institute of Post and Telecommunications (Rabat, Morocco). In 2010, he received the Ph.D degree in Networking and Computer Sciences with highest honors, jointly from University of Avignon (France) and the Mohammed V University (Rabat, Morocco).
He was a contractual Associate Professor at University of Avignon, from 2009 to 2012. Currently, he is a full-time Associate Professor at the National Higher School of Electricity and mechanics (ENSEM). In 2014, he obtained with highest honors the degree of Habilitation Qualification from Hassan II University of Casablanca.In July-September 2017, he has been a visiting Professor at Université du Québec à Montréal-Canada (UQAM).
Dr. Sabir is an active IEEE Senior member who serves as a reviewer for many prestigious international journals (IEEE, Springer, Elsevier, Wiley, etc.), and assumes various TPC Chair and other roles in many major international conferences (GC, ICC, WCNC, ICT, UNet, IWCMC, WIOPT, WF-IoT, 5G-WF, IEEE 5G Summit, WINCOM, etc.). He is being (has been) involved in several national and international/European projects. His current research interests include protocols design for 5G and B5G Networks, D2D-M2M-IoT and Infrastructure-less Networking, Tactile Internet, Ultra Reliable Low Latency communications,
Wireless Intelligence, Stochastic Learning, Networking Games, Business Models and Network Neutrality. He is a prolific researcher who has co-authored more than 120 journal articles, book chapters, and/or conference publications.
His work has been awarded in many events. His doctoral research has been funded by an excellence scholarship from INRIA-France (2007-2010). He also was a recipient of the graduate scholarship (2007-2010) from the Moroccan Centre for Scientific and Technical Research.
As an attempt to bridge the gap between academia and industry, Dr. Sabir founded the International Conference on Ubiquitous Networking (UNet, www.unet-conf.org), and co-founded the International Conference On Wireless Networks And Mobile Communications (WINCOM, www.wincom-conf.org).
\end{IEEEbiographynophoto}


\begin{IEEEbiographynophoto}{Mohamed Sadik}
is a Professor and the Chair of the Research and Cooperation
Department at the National Higher School of Electricity and Mechanics (ENSEM). He received his PhD in Electrical Engineering from the National Polytechnic Institute of Lorraine (INPL) (Nancy, France) in December 1992. Professor Sadik received his Master of Sciences and his Bachelor, both in Electronics, automatic and informatics, from University of Brest ( France). In 1993, Prof. Sadik joined the National Higher School of Electricity and mechanics; Hassan II University of Casablanca as an Assistant Professor of signal processing and other courses related
to computer sciences and wireless networking. In 2004, he became an Associate Professor at ENSEM. From 2008 to 2013, he served as a chair of Electrical Engineering Department at the same institution. His earlier research activities lied in the development of biomedical autonomous systems. Recently, he has been interested in embedded system, autonomous/smart systems applied to precision agriculture and environment. He is also working on security issues, in particular for
networking and embedded systems. He also participates (participated) in numerous R\&D Projects. Dr. Sadik is the head of the Networking, embedded Systems and Telecommunications (NEST) research group. He serves as TCP and reviewer for several international and national conferences including WOTIC 2011, ICT2013, NGNS 2014, etc. Dr. Sadik has been involved in the organization of numerous international conferences (IEEE ITC'13, IEEE NGNS'14, WINCOM'15, etc.). He also co- founded and chaired the international Symposium on Ubiquitous Networking 2015.
\end{IEEEbiographynophoto}

\begin{IEEEbiographynophoto}{Latif Ladid}
holds also the following positions: Emeritus Trustee, Internet Society,
Board Member IPv6 Ready \& Enabled Logos Program, and Board Member World
Summit Award. He is a Senior Researcher at the University of Luxembourg on multiple European Commission Next Generation Technologies IST Projects: 6INIT, www.6init.org – First Pioneer IPv6 Research Project; 6WINIT, www.6winit.org; Euro6IX, www.euro6ix.org; Eurov6, www.eurov6.org; NGNi, www.ngni.org; Project initiator SEINIT, www.seinit.org, and Project initiator SecurIST, www.securitytaskforce.org. He is also Board Member of 3GPP PCG
 (www.3gpp.org), 3GPP2 PCG (www.3gpp2.org), Vice Chair, IEEE ComSoc SDN-
NFV subC, member of UN Strategy Council, and member of the Future Internet Forum EU Member States (representing Luxembourg).
\end{IEEEbiographynophoto}



\end{document}